\documentclass[aps,prl,twocolumn,showpacs,floatfix,a4paper]{revtex4-2}
\usepackage{graphicx}
\usepackage[intlimits]{amsmath}
\usepackage{color}

\begin{document}

\title{Inducing a Metal-Insulator Transition through Systematic Alterations of Local Rewriting Rules in a Quantum Graph}
\author{Richard Berkovits}
\affiliation{Department of Physics, Jack and Pearl Resnick Institute, Bar-Ilan University, Ramat-Gan 52900, Israel}

\begin{abstract}
The Anderson localization transition in quantum graphs has garnered significant recent attention due to its relevance to many-body localization studies. Typically, graphs are constructed using top-down methods. Here, we explore a bottom-up approach, employing a simple local rewriting rule to construct the graph. Through the use of ratio statistics for the energy spectrum and Kullback-Leibler divergence correlations for the eigenstates, numerical analysis demonstrates that slight adjustments to the rewriting rule can induce a transition from a localized to an extended quantum phase. This extended state exhibits non-ergodic behavior, akin to the non-ergodic extended phase observed in the Porter-Rosenzweig model and suggested for many-body localization. Thus, by adapting straightforward local rewriting rules, it becomes feasible to assemble complex graphs from which desired global quantum phases emerge. This approach holds promise for numerical investigations and could be implemented in building optical realizations of complex networks using optical fibers and beam splitters.
\end{abstract}


\maketitle


Since its inception more than six decades ago \cite{anderson58}, Anderson localization has consistently captured the imagination of the physics community \cite{lee85, kramer93, ghur98, alhassid00, mirlin00, akkermans07, evers08}. Recently, much of this fascination has centered around the phenomenon of many-body localization (MBL) \cite{altshuler97, gornyi05, basko06, nandkishore15, alet18, abanin19}, where models of single-particle graphs have played a pivotal role in shaping its concepts \cite{altshuler97, gornyi05, basko06}. Subsequently, concerted efforts have been made to understand this phenomenon by constructing graph models that faithfully replicate the structure of couplings between non-interacting states in the Fock space of the microscopic many-body model. Consequently, substantial emphasis has been placed on the exploration of properties within single-particle graphs, with the primary goal of gaining a deeper understanding of MBL behavior.

Initially, these endeavors predominantly focused on Cayley tree structures \cite{deluca14, tikhonov16, sooner17, kravtsov18, biroli18, biroli22} that lack closed loops. However, recognition of the significance of closed loops has grown, leading to a shift in focus towards graph models incorporating such loops \cite{tikhonov16, garcia17, metz17, bera18, tikhonov19, tikhonov19a, garcia20, garcia20, tikhonov21, tarzia22, colmenarez22, garcia22, sierant23, kochergin23}. These graphs typically exhibit metallic and localized regimes dependent on the structural characteristics and level of disorder. These studies are in alignment with earlier research on Anderson localization within graph structures \cite{abou73, mirlin91, mirlin94, mirlin97, zhu00, sade03, sade05,jahnke08}.

Numerous graph types have been subject to study, encompassing a wide range, including but not limited to Cayley trees, random regular graphs, small-world graphs, and scale-free graphs. These graphs share a common characteristic: they are constructed in a top-down manner. This approach involves defining global properties, such as the average number of edges per node or their distribution, and subsequently employing an appropriate algorithm to shape the desired graph structure. Disorder is introduced through random connections between nodes, variations in on-site energies, or differences in the strength of the hopping matrix elements represented by the graph's edges. For most graphs, metallic and localized regimes are observed, corresponding to the effects of weak and strong disorder, as determined by an appropriate measure. Diverse behaviors in the transitions between metallic and localized states are observed across different graphs. In some cases, a classical quantum metal-insulator Anderson transition is evident, while in others, the transition is less pronounced, with a region displaying non-ergodic extended (NEE) behavior observed between the ergodic metallic and localized phases.

\begin{figure}
\includegraphics[width=9cm,height=!]{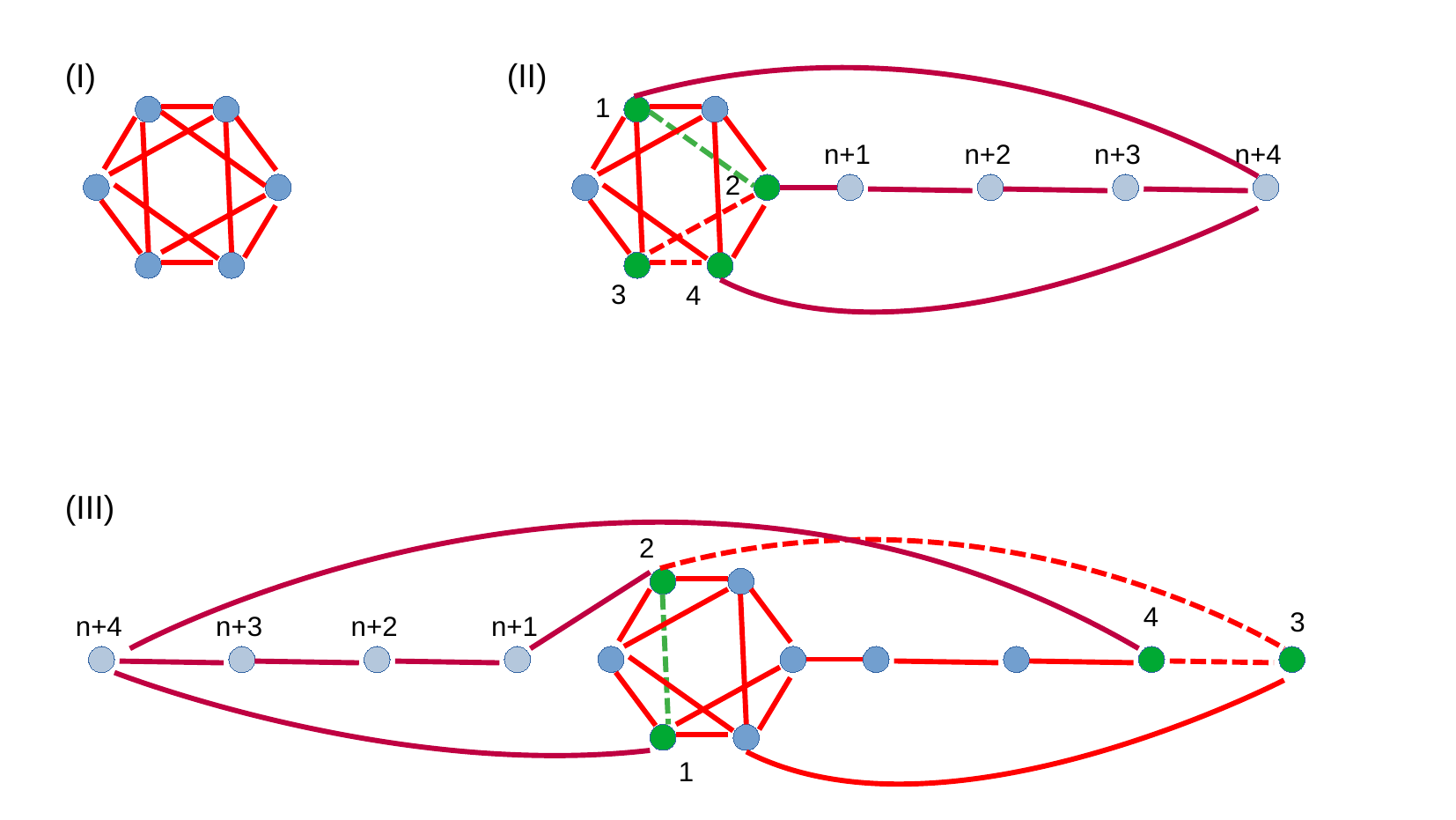}
\caption{\label{fig1}
  An illustration of two consecutive applications of the rewriting rule discussed in the text, where $d=2$ and $q=4$, starting from an initial seed of $N_0=6$ (depicted in (I)). (II) In the first iteration, a node (marked by the label $1$) is randomly chosen, and a random path along existing edges to nodes $2,3,4$ is selected (nodes are indicated by green circles and the edges connecting them by dashed lines). A chain of $4$ new nodes (light blue circles) is attached to node 2. New edges (maroon lines) are added between nodes $1, n+4$, $4, n+4$ and along the chain, while the edge between nodes $1,2$ is removed (green dashed line). (III) Second iteration of the rewriting rule.
}
\end{figure}

An alternative approach involves the bottom-up construction of a quantum graph. In this method, straightforward computational rules are employed to randomly select a motif within the graph (see Fig. \ref{fig1}), followed by the application of local modifications through the introduction of new nodes and the addition or rearrangement of edges in its vicinity. By iteratively applying this rule to an initial seed, a diverse array of random graphs can be generated and subjected to analysis. As will be demonstrated, continuous modifications to a selected local rewrite rule can drive the system through a metal-insulator transition, commencing from a localized phase, exhibiting a second-order transition into a NEE phase. This sequence bears resemblance to the first transition observed in the Rosenzweig-Porter model \cite{rosenzweig60, kravtsov15, monthus17, kravtsov18a, bogomolny18, nosov19, pino19, tomasi19, berkovits20, buijman22, tang22, altshuler23, venturelli23}.

Conceptually, this stochastic local rewrite approach draws inspiration from studies on cellular automata \cite{gardner70, wolfram83, schiff11}, tackling the question of whether a complex system with non-trivial behavior can emerge from simple local dynamical rules. Ref. \cite{wolfram02} extended these concepts to the construction of networks and graphs using straightforward assembly rules. Given the significant impact of factors like dimensionality, the prevalence of closed loops, coordination numbers, correlations, etc., on the quantum phases of graphs, the exploration of whether assembled graphs can be driven through different quantum phases by subtly adjusting the rewriting rules becomes an intriguing pursuit.

A graph is constructed following a specific type of local rewriting rule applied to a seed. For example a seed (Fig. \ref{fig1}(I)) of $N_0$ nodes, each connected by $4$ edges to the nearest and next nearest node. The following steps are reiterated until the graph has $N$ nodes: (i) Choose a node randomly (denoted as node $1$); (ii) Mark $d+1$ additional nodes (denoted $2, \ldots, d+2$ and represented by green circles and dashed edges in Fig. \ref{fig1}(II), where $d$ is a parameter) reachable from node $1$ by consecutive edges; (iii) Attach a chain of $q$ new nodes (denoted $n+1,n+2,\ldots n+q$, where $n$ is the current number of nodes) to node $2$ (the light blue nodes connected by maroon edges in Fig. \ref{fig1}(II)); (iv) Add edges between nodes $1, n+q$ and between $d+2, n+q$, and remove the edge between $1, 2$ (maroon and dashed green edges in Fig. \ref{fig1}(II)); (v) Reset the graph as the starting point for the next iteration. The subsequent iteration is depicted in Fig. \ref{fig1}(III). Repeating this procedure starting from the same initial seed, $M$ times will results in a ensemble of $M$ graphs whose properties are studied.

\begin{figure}
\includegraphics[width=9cm,height=!]{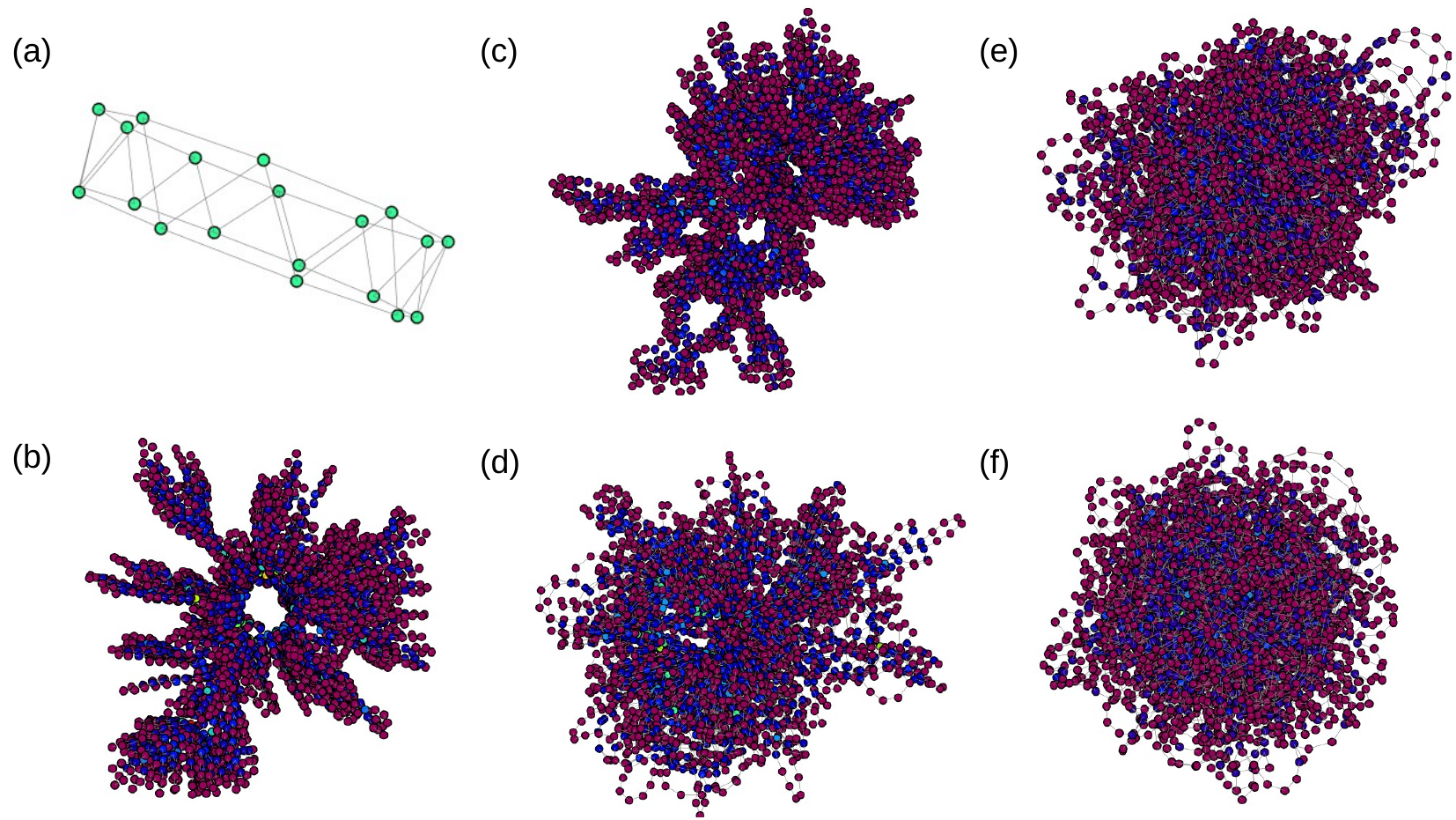}
\caption{\label{fig2}
Graphs constructed by the rewriting rule described in the text. (a) A typical seed with $N_0=24$. A representative graph for $q=4$ after reaching $N=4096$ for (b) $d=1$; (c) $d=2$; (d) $d=4$; (e) $d=6$; (f) $d=8$. The color of the nodes indicate the number of edges connected to node (red $2$, blue $3$, green $4$, cyan $5 \ldots$).
}
\end{figure}

The number of nodes, $d+2$, participating in stage (ii) of the algorithm and the number of nodes $q$ appearing in stage (iii) can be used as knobs to alter the overall structure of the graph emerging from the rewriting rule. As observed in Fig. \ref{fig2}, as $d$ is increased, the graph seems to become more compact. Our aim is to investigate whether changes of the local rewring rules by changing $q$ and $d$ are manifested in the spectral properties of these graphs. Since we intend to use $d$ as the control parameter for a given $q$, we need to define $d$ for non-integer values. In the case of $\lfloor d \rfloor < d < \lceil d \rceil$, stage (ii) is performed with $\lfloor d \rfloor + 2$ nodes with probability $\lceil d \rceil -d$ and with $\lceil d \rceil + 2$ nodes with probability $d - \lfloor d \rfloor$. Thus, $d$ can be tuned to any value larger than $1$ (since, for this rewriting rule, new edges are connected to $3$ nodes).

\begin{figure}
\includegraphics[width=10cm,height=!]{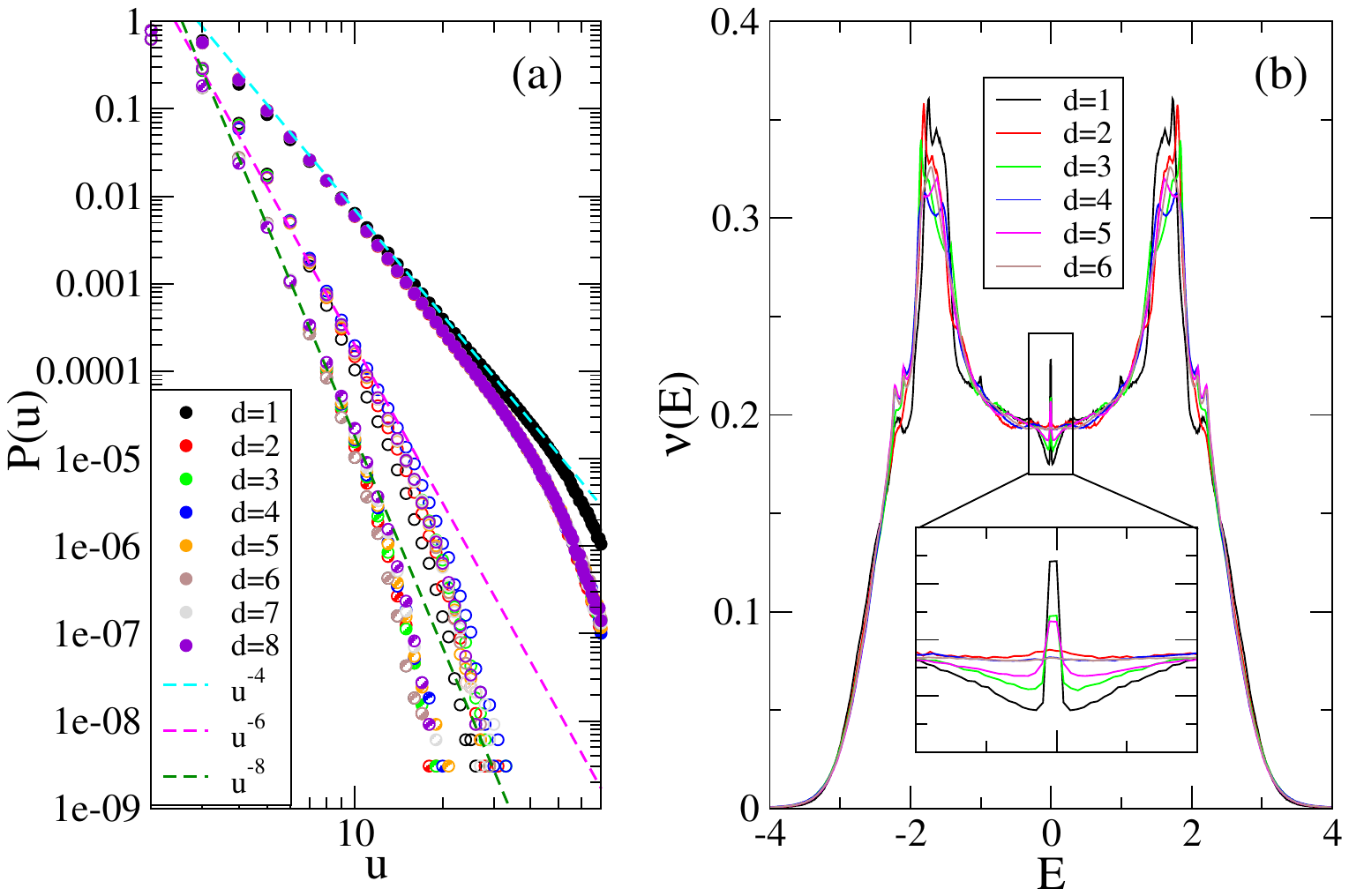}
\caption{\label{fig3}
(a) The probability that $u$ edges are connected to a single node, $P(u)$, for $q=1$ (filled symbols), $q=4$ (empty symbols) $q=8$ (line pattern symbols) and $d$ (see legend). The lines corresponds to $u^{-\lambda}$ with $\lambda=4,6,8$, (b) Density of states, $\nu(E)$, for $d=1,\ldots 6$. Zooming into the vicinity of the middle of the band, odd values of $d$ show a sharp peak at $E=0$. 
}
\end{figure}

To characterize the properties of the resulting graphs, one can examine the probability, denoted as $P(u)$, for a certain number of edges $u$ to be connected to a node. Naturally, $P(u)$ depends on the parameter $q$, as does the average number of edges per node, denoted as $\langle u \rangle$, which by construction yields $\langle u \rangle = 2 + 2/q$ for sufficiently large $N$. The distribution appears to adhere to a power law, represented as $P(u) \sim u^{-\lambda}$, where numerically, $\lambda(q,d)$ exhibits a strong dependence on $q$ and a weaker dependence on $d$. As depicted in Fig. \ref{fig3}a, $\lambda(q=1,d) \sim 4$, $\lambda(q=4,d) \sim 6$, and $\lambda(q=8,d) \sim 8$. Therefore, while the graphs display a power-law dependence, characteristic of scale-free graphs, the exponent $\lambda$ exceeds or equals four, suggesting a departure from scale-free features and instead aligning more closely with random graph behavior \cite{cohen03,voitalov19}.

The nodes and edges represent a quantum graph with the following Hamiltonian:
\begin{eqnarray} \label{hamiltonian}
H =
\displaystyle \sum_{i=1}^{N-1} \sum_{j>i}^{N} (t_{i,j}{\hat c}^{\dagger}_{i}{\hat c}_{j} + \text{h.c.}),
\end{eqnarray}
where $t_{i,j} = 0$ if there is no edge between nodes $i$ and $j$, while $t_{i,j} = \exp(-i \theta_{i,j})$ where $\theta_{i,j}$ is randomly distributed between $0$ and $2 \pi$ if an edge is present. The random phase associated with the hopping is added for several reasons. First, it smoothens the density of states (which will be discussed shortly). It also suppresses localization and provides a wider window for metallic behavior \cite{lee85}. Moreover, it provides a very natural experimental realization of the graphs discussed here where the edges are optical fibers of random length, while the nodes correspond to beam splitters \cite{jahnke08}. Ensembles of graphs of sizes $N=2^{10}-2^{15}$ are constructed for different values of $q$ and $d$ as described above, and the Hamiltonians are diagonalized using exact diagonalization to obtain the energy spectrum.

In Fig. \ref{fig3}b, the density of states as a function of energy, $\nu(E)$, is plotted for $q=4$ and various values of $d$. It can be observed that $\nu(E)$ exhibits only weak dependence on $d$ for $d\le 2$. For odd $d$'s, the density shows a residual sharp peak at $E=0$. These densities closely resemble those plotted in Ref. \cite{dorogovtsev03} (see Fig. 2 there) for complex scale-free graphs with a local tree-like structure. As noted in that reference, the overall shape of the density of states bears a striking resemblance to that of a quasi-one-dimensional system. The degeneracy at $E=0$ is also evident in other network models \cite{dorogovtsev03, aguiar05, yadav15} where $t_{i,j}=-1$.

The metal-insulator transition is evident in the energy spectrum of the system, marked by a transition in the nearest-neighbor level spacing distribution from Poisson statistics in the insulating phase to GUE (Gaussian Unitary Ensemble) statistics in the metallic phase \cite{shklovskii93}. A precise measure of the nearest-neighbor statistics is the ratio statistic \cite{oganesyan07, atas13}, defined as:
\begin{eqnarray} \label{ratio}
r_s &=& \langle \min (r_n,r_n^{-1}) \rangle, \\ \nonumber
r_s &=& \frac {E_n-E_{n-1}}{E_{n+1}-E_{n}},
\end{eqnarray}
here, $E_n$ represents the n-th eigenvalue of the graph, and $\langle \ldots \rangle$ denotes an average over different graphs in the ensemble for a given combination of $q$, $d$, and size $N$. This average is taken by considering a quarter of the eigenvalues centered around the $N/4$-th eigenvalue to circumvent the degeneracy at the center of the band and the tail region. For the Poisson distribution, the expected value is $r_s=2\ln(2) - 1 \approx 0.3863$, while for GUE, $r_s \approx 0.5996$ \cite{atas13}.

\begin{figure}
\includegraphics[width=10cm,height=!]{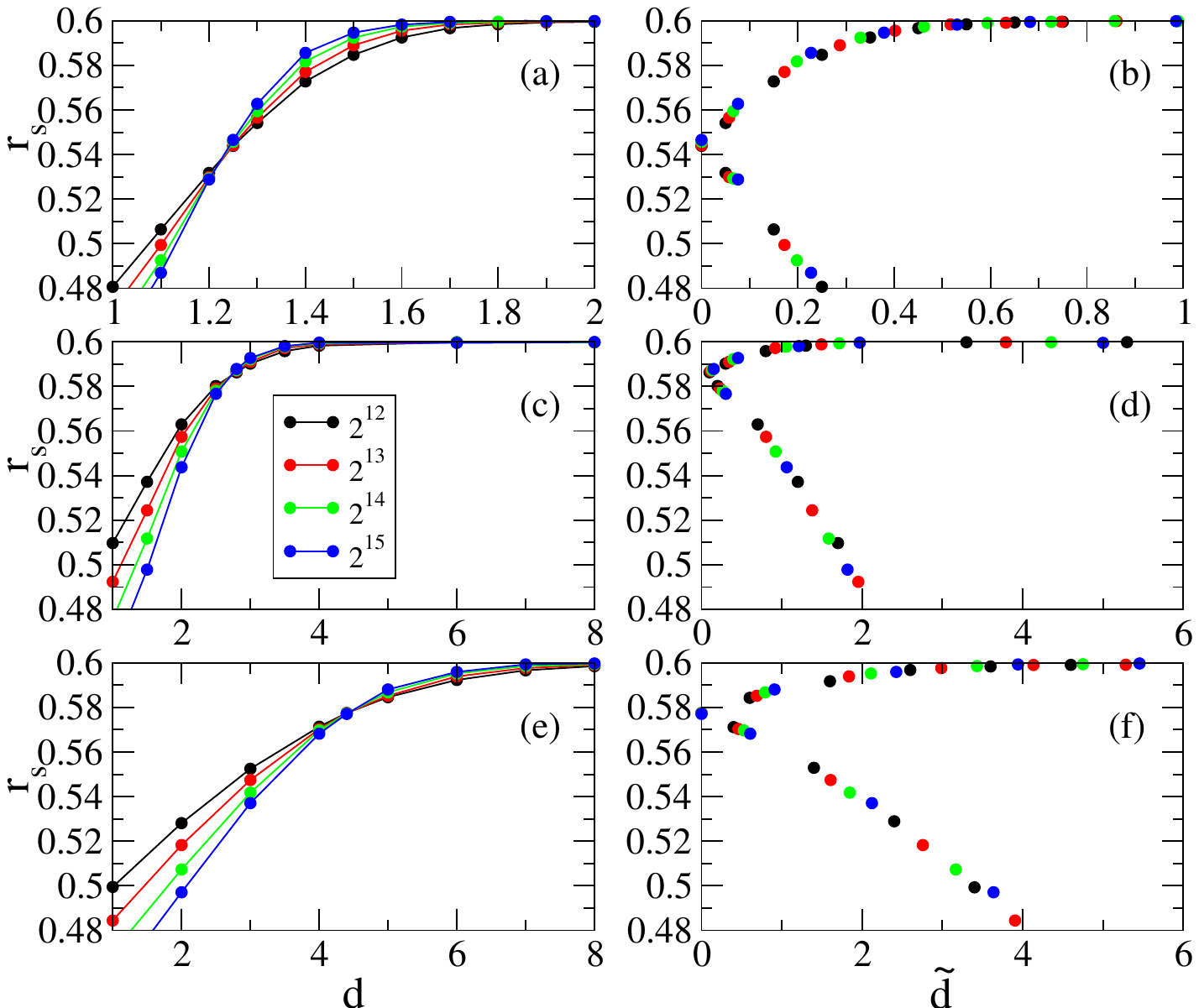}
\caption{\label{fig4}
Ratio statistics, $r_s$ as a function of $d$ for (a) $q=1$, (c) $q=4$, and (e) $q=8$, with an increasing number of nodes $N=2^{12},2^{13},2^{14},2^{15}$, exhibit the expected fanning of the curves in opposite directions from the critical point at $d_c(q=1)=1.25$, $d_c(q=4)=2.8$, and $d_c(q=8)=4.4$, as anticipated from a second-order phase transition. For large $d$, the GUE value of $r_s \sim 0.6$ becomes evident. Finite size scaling, $\tilde{d} = |d-d_c|(N/\tilde N)^{\beta}$, with $\tilde N=2^{12}$, is shown in (b,d,f). This demonstrates the expected finite size scaling behavior for an Anderson metal-insulator transition.
}
\end{figure}

Finite size scaling posits that in the insulating phase, $r_s$ approaches the Poisson value as $N$ grows, while in the metallic phase, $r_s$ approaches the GUE value. At the transition point $d_c$, $r_s$ should be independent of the graph size. Thus, plotting $r_s(d)$ for larger values of $N$ reveals more step-like curves, while the order of the curves switches. This is illustrated in Fig. \ref{fig4} (a,c,e), where $r(d)$ for different values of $N$ are plotted for (a) $q=1$, (c) $q=4$, and (e) $q=8$, displaying a clear crossing point at $d_c(q=1)=1.25$, $d_c(q=4)=2.8$, and $d_c(q=8)=4.4$. Another indicator of the second-order phase transition is finite size scaling, where the scaled values of $\tilde d$ should fall on two curves above and below the transition. Utilizing $\tilde{d} = |d-d_c|(N/\tilde N)^{\beta}$, with $\beta=0.2$, the appropriate $d_c(q)$, and $\tilde N=2^{12}$ chosen as the smallest graph size, Fig. \ref{fig4} (b,d,f) demonstrates the effectiveness of this scaling, confirming the metal-insulator transition.

Thus, the evidence for an insulator-metal transition at a critical value associated with the form of the local rewriting rule which constructs the graph appears solid. Nevertheless, as observed in the Rosenzweig-Porter model \cite{rosenzweig60, kravtsov15, monthus17, kravtsov18a, bogomolny18, nosov19, pino19, tomasi19, berkovits20, buijman22, tang22, altshuler23, venturelli23}, a transition in the ratio statistics may occur between a localized phase and an extended phase, which is not necessarily ergodic. To characterize the behavior of the extended phase, we will employ two different methods: singular value decomposition (SVD) of the energy spectrum \cite{fossion13,torres17,torres18,berkovits20}, and the Kullback-Leibler divergence (KL) correlation functions of the eigenstates \cite{kullback51,alet18,pino19,kravtsov20}.


For the SVD analysis, one compiles the $P$ eigenvalues considered for the calculation of $r_s$ from $M$ realizations of disorder into a matrix $X$ of size $M \times P$. Utilizing SVD, the matrix is decomposed as $X=U \Sigma V^T$, where $U$ and $V$ are $M\times M$ and $P \times P$ matrices, respectively, and $\Sigma$ is a diagonal matrix of size $M \times P$. The $r=\min(M,P)$ diagonal elements of $\Sigma$ represent the singular value amplitudes (SVA) $\sigma_k$ of $X$. The SVA are positive and can be ordered by size as $\sigma_1 \geq \sigma_2 \geq \ldots \sigma_r$. The behavior of the SVA squared ($\lambda_k=\sigma_k^2$) as a function of $k$ provides insights into the properties of the eigenvalues. Generally, the SVA follows a power law, $\lambda_k \sim k^{-\alpha}$, where $\alpha=1$ signifies the localized regime, while $\alpha=2$ characterizes the GUE phase \cite{fossion13,torres17,torres18}. For realistic physical models in the ergodic metallic regime, GOE/GUE behavior holds only up to the Thouless energy, leading to a transition to a different power law for small values of $k$ with $\alpha=1+D/2$, where $D$ represents the dimension \cite{berkovits21}. This power law exponent does not depend on disorder.

\begin{figure}
\includegraphics[width=10cm,height=!]{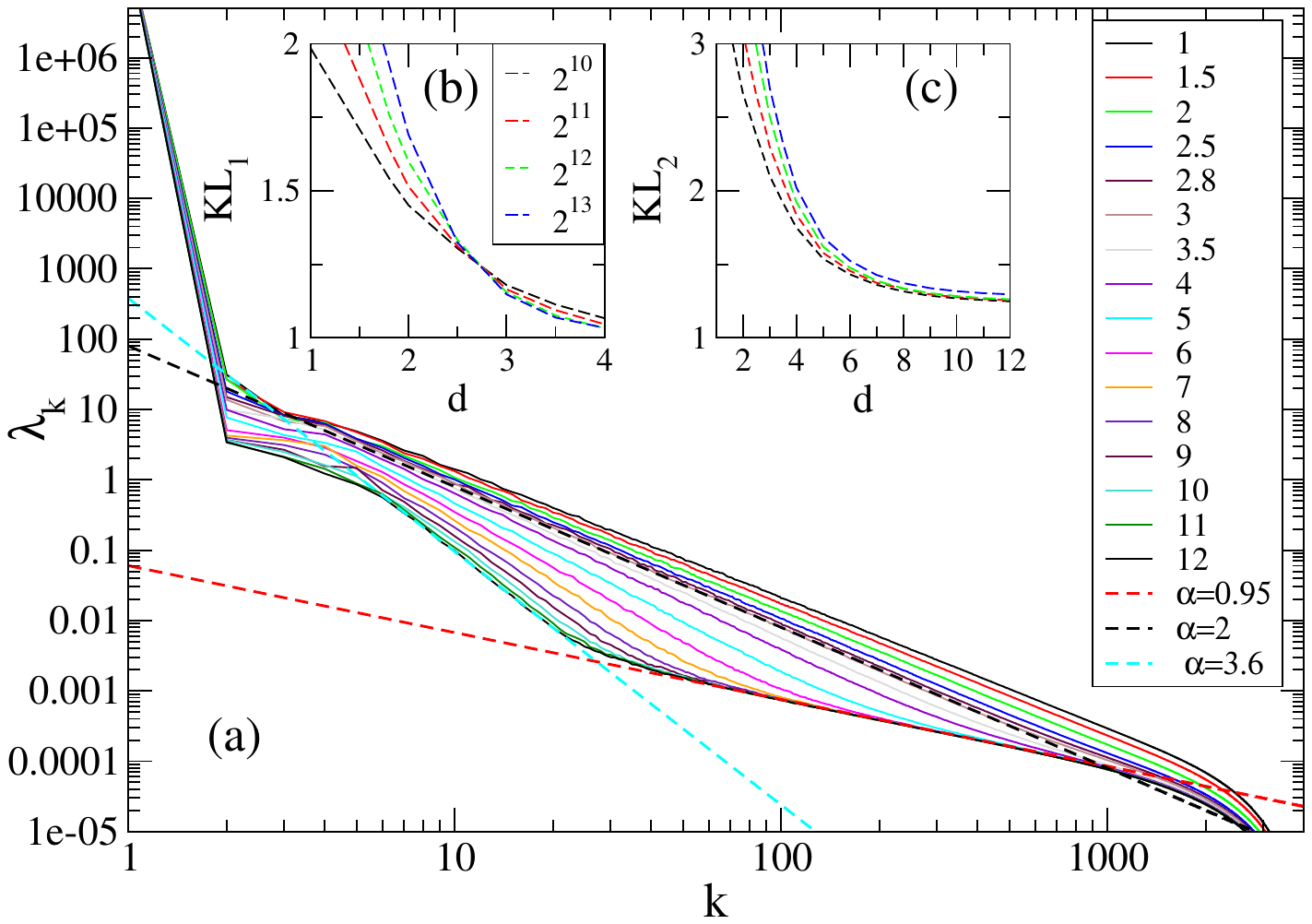}
\caption{\label{fig5}
  (a) The scree plot (SVA $\lambda_k$ vs. $k$) for $q=4$ with $M=4000$ realizations and $P=4000$ around the $N/4$ eigenvalue for graphs of size $2^{14}$ and different values of $d$. A power law $\lambda_k \sim k^{-\alpha}$ is expected for the Poisson statistics (with $\alpha=2$) and GUE statistics ($\alpha=1$). For $d>d_c(q=4)$, large $k$ follow a GUE power-law, while at lower $k$, a crossover to a steeper power occurs. $\alpha$ becomes larger as $d$ increases. (b) The Kullback-Leibler divergence between adjacent eigenstates in the same realization, $KL_1$, for $q=4$ and different values of $d$ for $M=1000$ realizations and $P=4000$ eigenvalues, and different sizes $d$ for different sizes $N=2^{10}, 2^{11}, 2^{12}, 2^{13}$. (c) Similar to (b), the Kullback-Leibler divergence between eigenstates of different realizations, $KL_2$, is presented.
}
\end{figure}

As demonstrated for the Rosenzweig-Porter model \cite{berkovits20}, in the NEE phase, the SVA for large $k$ (corresponding to small energy scales) follow metallic predictions ($\alpha=1$). For intermediate values of $k$, the power-law transitions to a different exponent. The closer one approaches the ergodic extended regime, this transition occurs at lower values of $k$, and $\alpha$ increases. Similar behavior is observed in Fig. \ref{fig5}a, where the SVA as a function of $k$ for different values of $d$ for $q=4$ and $N=10^{14}$ are plotted. For $d<d_c(q=4)=2.8$, a power law with $\alpha \sim 2$ is observed, as expected for the localized regime. For $d>d_c(q=4)$, large $k$ exhibit a power law with $\alpha \sim 1$. As $d$ increases, the crossover from $\alpha=1$ to a steeper power occurs at lower values of $k$, and $\alpha$ becomes larger, resembling the behavior in the NEE region of the Rosenzweig-Porter model \cite{berkovits20}.

A parallel corroboration for the NEE nature of this phase comes from examining the Kullback-Leibler correlation functions $KL_1$ and $KL_2$. By calculating the $p$-th eigenstates of the $m$-th realization, $|\psi^m_p\rangle$, in the nodes basis $|n\rangle$, it is possible to define $q^m_p(n)=|\langle n|\psi^m_p\rangle|^2$.  $KL_{1/2}$ may be written as:
\begin{eqnarray} \label{ratio}
  KL_1 &=& \sum^{M}_{m=1} \sum^{P}_{p=1} \sum^{N}_{n=1} q^m_p(n) \ln[q^m_p(n)/q^{m}_{p+1}(n)] /MP,
  \\ \nonumber
KL_2 &=& \sum^{M}_{m=1} \sum^{P}_{p=1} \sum^{N}_{n=1} q^m_p(n) \ln[q^m_p(n)/q^{m+1}_p(n)]/MP.
\end{eqnarray}
Thus, $KL_1$ probes the difference between adjacent eigenstates in the same realization, while $KL_2$ probes the difference between eigenstates in different realizations. In the ergodic extended phase, $KL_{1/2} \sim O(1)$. On the other hand, for the localized phase, the centers of two different eigenstates are at different uncorrelated nodes, and therefore $KL_{1/2} \sim \ln(N)$. Therefore, for a phase transition between the localized regime and the ergodic extended phase, one expects $KL_1$ and $KL_2$ to switch behavior from increasing as a function of $N$ to approaching a constant value \cite{pino19}.

Indeed, as shown in Fig. \ref{fig5}b, $KL_1$ exhibits such behavior with a typical critical point at $d=d_c(q=4)=2.8$, while for large $d$, $KL_1\rightarrow 1$. One characteristic property of an NEE phase is that eigenstates don't cover all nodes, and adjacent eigenstates in the same realization tend to be strongly overlapping \cite{kravtsov20}. This implies that $KL_1$ is not strongly affected and shows similar behavior to the extended ergodic behavior, while $KL_2$, for which the two eigenstates occupy different realizations, is more akin to the localized regime. Thus, it is not expected to be strongly influenced by the transition from the localized to the NEE phase. Indeed, as depicted in Fig. \ref{fig5}c, there is no change in behavior as $d_c(q=4)$ is traversed, strengthening the case that the transition is between a localized and an NEE phase.

Hence, complex graphs emerge from simple local rewrite rules, with the desired quantum phase manifesting itself. It's intriguing to speculate whether further fine-tuning of the rewrite rule could unveil even more intricate phase landscapes, potentially spanning from localized to non-ergodic extended (NEE) to ergodic extended phases. Not only does this approach offer clear numerical advantages, but it also serves as a blueprint for constructing quantum graphs in the lab using basic components like optical fibers and beam splitters, employing straightforward assembly protocols.


\begin{acknowledgments}
Useful discussions with I. Bonamassa and S. Havlin are gratefully acknowledged. I thank N. Krombine for writing the graphical software used in Fig. \ref{fig2}.
\end{acknowledgments}


\end{document}